\documentclass[aps,prd,showpacs,nofootinbib,twocolumn,floatfix,superscriptaddress,preprintnumbers]{revtex4}
\usepackage{amsmath}
\usepackage{amssymb}
\usepackage{epsfig}
\usepackage{graphicx}
\usepackage{stmaryrd}

\usepackage{color}

\newcommand{\nn}{\nonumber}

\def\10{$SO(10)$}
\def\21{SU(2) $\otimes$ U(1) }

\def\422{$SU(4) \otimes SU(2) \otimes SU(2)$}
\def\321{SU(3) $\otimes$ SU(2) $\otimes$ U(1)}

\newcommand {\ignore}[1]{}

\def\lsim{\raise0.3ex\hbox{$\;<$\kern-0.75em\raise-1.1ex\hbox{$\sim\;$}}}
\def\gsim{\raise0.3ex\hbox{$\;>$\kern-0.75em\raise-1.1ex\hbox{$\sim\;$}}}
\def\vev#1{\left\langle #1\right\rangle}
\newcommand{\AddrAHEP}{%
  AHEP Group, Institut de F\'{\i}sica Corpuscular --
  C.S.I.C./Universitat de Val{\`e}ncia \\
  Edificio Institutos de Paterna, Apt 22085, E--46071 Valencia, Spain}

 \renewcommand{\baselinestretch}{1.33}
 \newcommand{\ba}{\begin{array}}
\newcommand{\ea}{\end{array}}
\relax
\def\321{$SU(3)\times SU(2)\times U(1)$}

\newcommand{\Sol}  {\textrm{sol}}

\newcommand{\Atm}  {\textrm{atm}}

\newcommand{\Dms}  {\Delta m^2_\Sol}
\newcommand{\Dma}  {\Delta m^2_\Atm}

\begin{document}
\preprint{IFIC/09-19}
\renewcommand{\Huge}{\Large}
\renewcommand{\LARGE}{\Large}
\renewcommand{\Large}{\large}

\title{\bf A4-based tri-bimaximal mixing within inverse and linear
  seesaw schemes}
\author{M.~Hirsch} \email{mahirsch@ific.uv.es} \affiliation{\AddrAHEP}
\author{S.~Morisi} \email{morisi@ific.uv.es} \affiliation{\AddrAHEP}
\author{J.~W.~F.~Valle} \email{valle@ific.uv.es} \affiliation{\AddrAHEP}

\date{\today}

\begin{abstract}

  We consider tri-bimaximal lepton mixing within low-scale seesaw
  schemes where light neutrino masses arise from TeV scale physics,
  potentially accessible at the Large Hadron Collider (LHC).
  Two examples are considered, based on the $A_4$ flavor symmetry
  realized within the inverse or the linear seesaw mechanisms. Both
  are highly predictive so that in both the light neutrino sector
  effectively depends only on three mass parameters and one Majorana
  phase, with no CP violation in neutrino oscillations.  
  We find that the linear seesaw leads to a lower bound for neutrinoless
  double beta decay while the inverse seesaw does not. 
  The models also lead to potentially sizeable decay rates for lepton
  flavor violating processes, tightly related by the assumed flavor
  symmetry. 

\end{abstract}
\maketitle

\section{Introduction}

Neutrino mass generation in the Standard Model is likely to come from
a basic dimension-five operator that violates lepton
number~\cite{Weinberg:1980bf}.  Little is known about the ultimate
origin of this operator, including the nature of the underlying
mechanism, its characteristic scale and/or flavor
structure. Correspondingly, it has many possible realizations
involving the exchange of scalar and/or fermions at the tree and/or
radiative level~\cite{Valle:2006vb}.

In a broad class of models the exchange of heavy gauge singlet
fermions induces neutrino masses via what is now called type-I
seesaw~\cite{gell-mann:1980vs,yanagida:1979,mohapatra:1981yp,schechter:1980gr,Lazarides:1980nt}.
An attractive mechanism called inverse seesaw has long been proposed
as an alternative to the simplest type-I
seesaw~\cite{Mohapatra:1986bd} (for other extended seesaw schemes see,
e.g.~\cite{Wyler:1983dd,Akhmedov:1995vm,Barr:2005ss}).
In addition to the left-handed SM neutrinos $\nu$ in the inverse
seesaw model ones introduces two \321 singlets $\nu^c, \, S$.  In the
basis $\nu,\,\nu^c, \, S$ the effective neutrino mass matrix is
\begin{equation}
M_\nu=\left(
\begin{array}{ccc}
0&M_D&0\\
M_D^T &0&M\\
0&M^T&0
\end{array}
\right),
\end{equation}
that can be simply justified by assuming a $U(1)_L$ global lepton
number symmetry.  Neutrinos get masses only when $U(1)_L$ is
broken. The latter can be arranged to take place at a low scale, for
example through the $\mu SS$ mass term in the mass matrix given below,
\begin{equation}\label{minv}
M_\nu=\left(
\begin{array}{ccc}
0&M_D&0\\
M_D^T&0&M\\
0&M^T&\mu
\end{array}
\right),
\end{equation}
After $U(1)_L$ breaking the effective light neutrino mass matrix
is given by
\begin{equation}\label{inv}
M_\nu=M_DM^{T^{-1}}\mu M^{-1}M_D^T.
\end{equation}
so that, when $\mu$ is small, $M_\nu$ is also small, even when $M$
lies at the electroweak or TeV scale.  In other words, the smallness
of neutrino masses follows naturally since as $\mu \to 0$ the lepton
number becomes a good symmetry~\cite{'tHooft:1979bh} without need for
superheavy physics.

The smallness of the parameter $\mu$ may also arise dynamically in
sypersymmetric models and/or spontaneously in a Majoron-like scheme
with $\mu\sim \vev{\sigma}$ where $\sigma$ is a \321 singlet
\cite{Gonzalez-Garcia:1988rw}. In the latter case, for sufficiently
low values of $\vev{\sigma}$ there may be Majoron emission effects in
neutrinoless double beta decay~\cite{berezhiani:1992cd}.

Recently another alternative seesaw scheme called linear seesaw has
been suggested from $SO(10)$~\cite{Malinsky:2005bi}. Here we consider
a simpler variant of this model based just on the framework of the
\321 gauge structure. In the basis $\nu,\,\nu^c, \, S$ the effective
neutrino mass matrix is
\begin{equation}\label{mlin}
M_\nu=\left(
\begin{array}{ccc}
0&M_D&M_L\\
M_D^T&0&M\\
M_L^T&M^T&0
\end{array}
\right).
\end{equation}
Here the lepton number is broken by the  $M_L\,\nu S$ term, 
and the effective light neutrino mass is given by
\begin{equation}\label{lin}
M_\nu=M_D(M_L M^{-1})^T+(M_L M^{-1}){M_D}^T.
\end{equation}

In addition to indications of non-vanishing neutrino mass, neutrino
oscillation
experiments~\cite{fukuda:2002pe,ahmad:2002jz,araki:2004mb,Kajita:2004ga,ahn:2002up}
indicate a puzzling structure~\cite{Maltoni:2004ei} of the elements of
the lepton mixing matrix, at variance with the quark mixing angles.

In this paper we consider the possibility of predicting lepton mixing
angles from first principles, in the framework of the inverse or
linear seesaw mechanisms to generate light neutrino masses.  An
attractive phenomenological ansatz for leptons
mixing~\cite{Harrison:2002er} is the tribimaximal (TBM) one
\begin{equation}
\label{eq:HPS}
U_{\textrm{HPS}} = 
\left(\begin{array}{ccc}
\sqrt{2/3} & 1/\sqrt{3} & 0\\
-1/\sqrt{6} & 1/\sqrt{3} & -1/\sqrt{2}\\
-1/\sqrt{6} & 1/\sqrt{3} & 1/\sqrt{2}
\end{array}\right)
\end{equation}
which is equivalent to the following values for the lepton mixing
angles: $\tan^2\theta_{\Atm}=1$, $\sin^2\theta_{\textrm{Chooz}}=0$ and
$\tan^2\theta_{\Sol}=0.5$, providing a good first approximation to the
values indicated by current neutrino oscillation data.

Below we give two simple $A_4$ flavor symmetry realizations of the TBM
lepton mixing pattern within the above seesaw schemes.
For example, for the inverse seesaw case possible schemes are
summarized in Table~\ref{tab0}.


\begin{table}[h!]
\begin{center}
\begin{tabular}{|l|ccc|ccc|c|}
\hline
cases & $1)$& $2)$&$3)$ &$4)$ &$5)$ & $6)$&$7)$  \\
\hline
$M_D$  & $M_0$&$I$&$I$&$I$ & $M_{0}$& $M_{0}$&$M_{0}$ \\
\hline
$M$  &$I$&$M_0$&$I$&$M_{0}$ & $I$&$M_{0}$  & $M_{0}$\\
\hline
$\mu$  & $I$& $I$&$M_0$&$M_{0}$& $M_{0}$&$I$ & $M_{0}$\\
\hline
\end{tabular}\caption{Possible TBM inverse seesaw schemes.}\label{tab0}
\end{center}
\end{table}


Recall that $A_4$ is the group of the even permutations of four
objects.  Such a symmetry was introduced to yield
$\tan^2\theta_{\Atm}=1$ and $\sin^2\theta_{\textrm{Chooz}}=0$
\cite{Ma:2001dn,babu:2002dz}.  Most recently $A_4$ has also been used
to derive $\tan^2\theta_{\Sol}=0.5$ \cite{Altarelli:2005yp}.  The
group $A_4$ has 12 elements and is isomorphic to the group of the
symmetries of the tetrahedron, with four irreducible representations,
three distinct singlets $1$, $1'$ and $1''$ and one triplet $3$. For
their multiplications see for instance Ref.~\cite{Altarelli:2005yp}.

If the charged lepton matrix $M_l$ is diagonalized on the left by the
magic matrix $U_{\omega}$
\begin{equation}\label{Uom}
U_\omega=\frac{1}{\sqrt{3}}\left(
\begin{array}{ccc}
1&1&1\\
1&\omega&\omega^2\\
1&\omega^2&\omega
\end{array}
\right),
\end{equation}
(with $\omega\equiv \exp{i\pi/3}$) we have tri-bimaximal mixing if the
light neutrino mass matrix has the structure
\begin{equation}
M_{0}=\left(
\begin{array}{ccc}
A&0&0\\
0&B&C\\
0&C&B
\end{array}
\right).
\end{equation}

We note that $M_{0}^{-1}$ has the same structure as $M_{0}$.  This
implies that, taking any one (or more) of the three $M_D, M, \mu $
matrices as having the $M_{0}$ structure, with the remaining ones
proportional to the identity matrix $I$ one obtains a light neutrino
mass matrix of TBM-type, namely
\begin{equation}\label{tbmt}
\left(
\begin{array}{ccc}
x&y&y\\
y&x+z&y-z\\
y&y-z&x+z
\end{array}
\right),
\end{equation}
leading to many potential ways to obtain the TBM mixing pattern within
an inverse seesaw mechanism.  In table~\ref{tab0} we list all possible
tri-bimaximal schemes.

\section{Tri-bimaximal inverse seesaw }

As illustrative example we consider the case with
\begin{equation}\label{ass1}
M_D\propto M_{0},\qquad M\propto I,\qquad \mu \propto I
\end{equation}
Below we will give a flavor model for such a case.  When we go to the
basis where charged leptons are diagonal Eq.~(\ref{Uom}), we have
$M_D\propto U_\omega M_{0},\, M\propto I,\, \mu \propto I$.

The light neutrino mass matrix arises from the inverse seesaw relation in
eq.\,(\ref{inv}) and we have
\begin{equation}\label{mnu1}
M_\nu\sim U_\omega M_{0}M_{0}^T U_\omega^T
\end{equation}
which is of TBM-type (\ref{tbmt}).
For example, for the case of real $M_0$ we have only three mass
parameters in the model, two of which are determined by neutrino
oscillations~\cite{Maltoni:2004ei} and the third is related to the
overall scale of neutrino mass that can be probed in tritium and
double beta decays. In the general case one can see that there is no
CP violation in neutrino oscillations, so that only a Majorana phase
survives. This is in sharp contrast with the generic form of the
inverse seesaw, which has CP violation even in the massless neutrino
limit~\cite{Branco:1989bn}.

The matter fields are assigned  as in table \ref{tab2}.\\
\begin{table}[h!]
\begin{center}
\begin{tabular}{|c|c|c|c|c||c|c|c|}
\hline
&$L$ & $l^c$ & $\nu^c$& $S$ & $h$ & $\xi, \phi$& $\xi'\phi'$ \\
\hline
$SU_L(2)$ & 2& 1& 1& 1&2 & 2&2\\
$Z_3$&$\omega$ &  $\omega$&$1$ &1 & 1  &  $\omega^2$&$\omega$\\
$A_4$&3&3&3&3&1&1,3&1,3\\
\hline
\end{tabular}\caption{Matter assignment for inverse seesaw model.}\label{tab2}
\end{center}
\end{table}

The renormalizable~\footnote{Here we have introduced several Higgs
  doublets. We can equivalently avoid having many Higgs doublets by
  introducing corresponding scalar electroweak singlet {\it flavon}
  fields.}  Lagrangian invariant under the symmetry $A_4\times Z_3$ is

\begin{equation}
\begin{array}{l}
\mathcal{L} = {Y_D}^k_{ij} \, L_i \, \nu^c_j (\phi_k+ \xi )+M_{ij}\,  \nu^c_i S_j + \\
\mu_{ij} S_{i} S_{j} + Y_{l_{ij}}^k   L_i\, l^c_j (\phi'_k+ \xi' )
\end{array}
\end{equation}
where from $A_4$-contractions we have that the couplings are given in
Eq.~(\ref{ass1}), $\mu= v_\mu I$, $M=v_M I$.
However when $\xi$ takes a
vacuum expectation value (vev) and
\begin{equation}\label{phi1}
\vev{\phi} \sim (1,0,0).
\end{equation}
we have in general
\begin{equation}\label{DA4}
M_D=\left(
\begin{array}{ccc}
a&0&0\\
0&a&b_1\\
0&b_2&a
\end{array}
\right).
\end{equation}
In contrast to $M_{0}$ such a matrix is not symmetric. Here we assume
the {\it ad hoc} relation $b_1=b_2=b$. Such a relation can be obtained
in the context of an $SO(10)$ model or by assuming $S_4$ flavor symmetry
instead of $A_4$ \footnote{The reason is that in $A_4$ $3\times 3
  =1+1'+1''+3_S+3_A$ and we must take also the antisymmetric
  contraction for Dirac mass terms.  $S_4$ is the group of permutation
  of four objects and $3\times 3 =1+2+3_1+3_2$ where $3_1$ and $3_2$
  are distinct irreducible representations.}.

The light neutrino mass eigenvalues are
\begin{equation}\label{eigiss}
\{m_1,m_2,m_3\}=\frac{v_\mu}{v_M^2}
\{(a+b)^2,a^2,-(a-b)^2\}.
\end{equation}
When also $\xi'$ takes a vev along
\begin{equation}\label{phi2}
\langle \phi' \rangle\sim (1,1,1).
\end{equation}
we have 
\begin{equation}
M_l=\left(
\begin{array}{ccc}
\alpha&\beta&\gamma\\
\gamma&\alpha&\beta\\
\beta&\gamma&\alpha
\end{array}
\right)=
U_\omega\left(
\begin{array}{ccc}
m_e&0&0\\
0&m_\nu & 0 \\
0 & 0&m_\tau
\end{array}
\right)U_\omega^\dagger.
\end{equation}
Therefore the charged lepton mass matrix is diagonalized on the left
by the magic matrix $U_\omega$ as required.

We note that when the Higgs doublets $\phi$ and $\phi'$ take nonzero
vevs, the $A_4$ symmetry breaks spontaneously into its two subgroups,
namely $Z_2$ and $Z_3$, respectively. The consequence of such a {\it
  misalignment} is to have a large mixing in the neutrino sector. The
problem how to get such a misalignment has been studied in many
contexts \cite{alg}.

\section{Tri-bimaximal linear seesaw }

We now consider the case of the linear seesaw, see eqs.\,(~\ref{mlin}) and
(\ref{lin}). As for the inverse seesaw, there are different possible
choices for $M_D,M, M_L$ that can lead to the TBM structure.  
We take as example the case with 
\begin{equation}\label{ass2}
M_D \propto M_{0},\qquad M \propto I,\qquad M_L \propto I.
\end{equation}
When we go to the basis where charged leptons are diagonal
(\ref{Uom}), we have
$M_D \propto U_\omega M_{0},\, M \propto I,\, M_L \propto U_\omega$.

From eq.\,(\ref{lin}) the light neutrinos mass matrix is given as
\begin{equation}\label{mnu2}
M_\nu\sim  U_\omega M_{0}U_\omega^T+U_\omega M_{0}^TU_\omega^T.
\end{equation}
We note that in contrast to the inverse seesaw, in the linear seesaw
case the light neutrino mass matrix $M_\nu$ in eq.\,(\ref{mnu2}) is of
TBM type also when $M_{0}$ is given by eq.\,(\ref{DA4}) without any
{\it ad hoc} symmetry assumption.
Again, as before, we note that for the case of real $M_0$ there are
only three mass parameters, two of which can be traded by the neutrino
oscillation mass splittings~\cite{Maltoni:2004ei}, with the remaining
one fixing the overall neutrino mass scale. Even in the presence of
complex phases in $M_0$ there is no CP violation in neutrino
oscillations, and only a Majorana phase remains (see below).

As an illustrative example we describe a model based on $A_4$ flavor
symmetry, in table \ref{tab3}.
\begin{table}[h!]
\begin{center}
\begin{tabular}{|c|c|c|c|c||c|c|c|}
\hline
&$L$ & $l^c$ & $\nu^c$& $S$ & $h$ & $\xi, \phi$& $\xi'\phi'$ \\
\hline
$SU_L(2)$ & 2& 1& 1& 1&2 & 2&2\\
$Z_3$&$\omega$ &  $1$&$\omega$ &$\omega^2$ & 1  &  $\omega$&$\omega^2$\\
$A_4$&3&3&3&3&1&1,3&1,3\\
\hline
\end{tabular}\caption{Matter assignment for   linear seesaw model.}\label{tab3}
\end{center}
\end{table}

The invariant Lagrangian is 
\begin{equation}
\begin{array}{l}
\mathcal{L} = {Y_D}^k_{ij} \, L_i\, \nu^c_j (\phi_k+ \xi )+M_{ij}\,  \nu^c_i S_j +Y_{L_{ij}} L_{i}h\, S_{j}\\
+ Y_{l_{ij}}^k   L_i\, l^c_j (\phi'_k+ \xi' )
\end{array}
\end{equation}
where the couplings are given as in Eq.~({\ref{ass2}}).

After the scalar fields take vevs obeying the alignment conditions
given in eqs.\,(\ref{phi1}) and (\ref{phi2}), the resulting Yukawa
couplings are given by (\ref{ass2}) and therefore the light neutrino
mass matrix is diagonalized by TBM in the basis where charged leptons
are diagonal as explained above.

The light neutrino eigenvalues are given by
\begin{equation}\label{eiglin}
\{m_1,m_2,m_3\}=\frac{v_L}{v_M}
\{(a+b),a,-(a-b)\}.
\end{equation}

\section{phenomenology}

Above we have introduced two very simple models based respectively on
inverse and linear seesaw mechanisms. 
Due to the assumed flavor symmetry they are highly restrictive. By
construction, the lepton mixing matrix in both models is predicted to
be tribimaximal and neutrino phenomenology is effectively described by
just three mass parameters and a phase.
Two of them are the neutrino squared-mass splittings well-determined
in neutrino oscillations.  The other mass parameter characterizes the
absolute neutrino mass scale which will be probled in tritium and
neutrinoless double beta decay searches, as well as cosmology.

As we have noted already, there is no CP violation in neutrino
oscillations, and only one Majorana phase remains and affects the
predictions for neutrinoless double beta decay (see below).

\textbf{Neutrinoless double beta decay}

\begin{figure}[!h]
\begin{center}
\includegraphics[angle=0,height=6.5cm,width=0.45\textwidth]{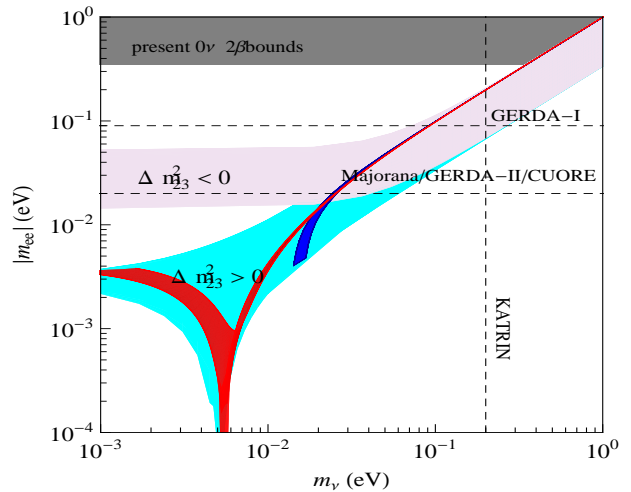}
\caption{Neutrinoless double beta decay parameter $m_{ee}$ as a
  function of the lightest neutrino mass for inverse seesaw (red) and
  linear seesaw (blue).  The cyan and purple bands represent
  respectively the generic regions 
  allowed by current data with lepton mixing angles fixed to be the
  tri-bimaximal values. For references to experiments see \cite{0nubb}.}
\label{fig:dbd}
\end{center}
\end{figure}

Despite their similarity, one can distinguish these models
phenomenologically since eqs\,(\ref{inv}) and (\ref{lin}) give rise to
different neutrino mass spectra and this implies different
expectations for neutrinoless double beta decay, as illustrated in
Fig.~\ref{fig:dbd} (similar predictions have been made within
A4-symmetric type-I seesaw models, as shown, for example in
Ref.~\cite{Hirsch:2008rp}).

One sees that, in contrast to the inverse seesaw, in the linear seesaw
case there is a lower bound on the neutrinoless double beta decay rate
despite the fact that we have a normal neutrino mass hierarchy.
In contrast, the effect of the Majorana phase in the inverse seesaw
can cause full cancellation in the decay rate.

\textbf{Lepton flavor violating decays}

In the inverse and linear seesaw models studied here, the neutrino
mass matrix is a $9\times 9$ symmetric matrix, see eqs\,(\ref{minv})
and (\ref{mlin}). This is diagonalized by a corresponding unitary
matrix $U_{\alpha\beta}$ of the same dimension, $\alpha,~\beta=1...9$,
leading to three light Majorana eigenstates $\nu_{i}$ with $i=1,2,3$
and six heavy ones $N_j$ with $j=4,..,9$.
The effective charged current weak interaction is characterized by a
rectangular lepton mixing matrix
$K_{i\alpha}$~\cite{schechter:1980gr}.
\begin{equation}
\mathcal{L}_{CC}=\frac{g}{\sqrt{2}}K_{i\alpha}\overline{L}_i\gamma_\mu (1+\gamma_5) N_\alpha \, 
W^{\mu}. 
\end{equation}
where $i=1,2,3$ denote the left-handed charged leptons and $\alpha$
the neutrals.  The contribution to the decay \(l_i \to l_j \gamma\)
arises at one loop (see for instance
\cite{Bernabeu:1987gr,Deppisch:2004fa}) from the exchanges of the six
heavy right-handed Majorana neutrinos $N_j$ which couple subdominantly
to the charged leptons.

The well-known one loop contribution to this branching ratio is
given by \cite{Ilakovac:1994kj} 
\begin{equation}
Br(l_i\to l_j \gamma)= \frac{\alpha^3s_W^2}{256 \pi^2}\frac{m_{l_i}^5}{M_W^4}
\frac{1}{\Gamma_{l_{i}}}|G_{ij}|^2
\end{equation}
where
\begin{equation}\label{def:G}
\begin{array}{l}
G_{ij}=\sum_{k=4}^9 K^*_{ik} K_{jk} G_\gamma\left(\frac{m^2_{N_k}}{M_W^2}\right)\\
G_\gamma(x)=-\frac{2 x^3+5 x^2 -x}{4 (1-x^3)}-\frac{3 x^3}{2(1-x)^4}\ln x
\end{array}
\end{equation}
We note that, thanks to the admixture of the TeV states in the charged
current weak interaction, this branching ratio can be sizeable even in
the absence of supersymmetry~\cite{Bernabeu:1987gr}. Similar results
hold for a class of LFV processes, including nuclear mu-e
conversion~\cite{Deppisch:2005zm}. As already noted, the rates for
mu-e conversion and \(\mu \to e \gamma\) are strongly correlated in
this model. These are the most stringently constrained LFV decays.

\begin{figure}[!h]
\begin{center}
\includegraphics[angle=0,height=6cm,width=0.45\textwidth]{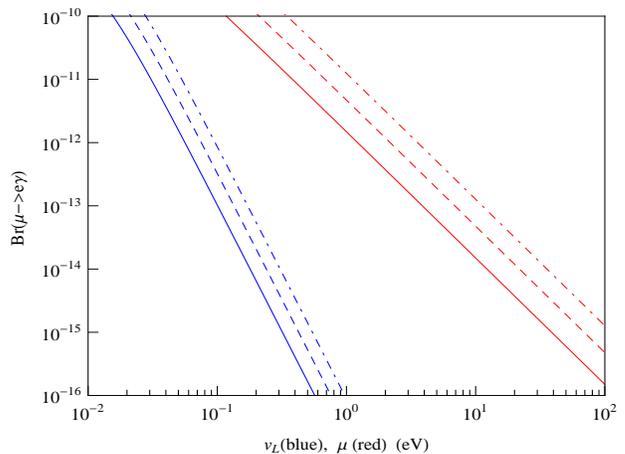}
\caption{$Br(\mu\to e \gamma)$ versus the lepton number violation
  scale: $\mu$ for the inverse seesaw (red color), and $v_L$ for the
  linear seesaw (blue color). In both cases, $M$ is fixed as $M=100 \,
  GeV$ (continous line), $M=200\, GeV$ (dashed line) and $M=1000\,
  GeV$ (dot-dashed line).}
\label{fig:MEGBR}
\end{center}
\end{figure}

The simplicity of their mass matrices, which are expressed in terms of
very few parameters, makes the current models especially restrictive and
this has an impact in the expected pattern of LFV decays.
In contrast to the general case considered in
\cite{Deppisch:2004fa,Deppisch:2005zm}, here we can easily display the
dependence of the $\mu \to e\gamma$ branching ratio on the new physics
scale represented by the parameters $M \sim$ TeV and the parameters
$\mu$ or $v_L$ characterizing the low-scale violation of lepton
number, since both are simply proportional to the identity matrix in
flavor space. This is illustrated in Fig.~\ref{fig:MEGBR}.

Note also that, in contrast to a generic inverse or linear seesaw
model, in our $A_4$ based models the structure of the matrix $G_{ij}$
is completely fixed, and this leads to predictions for ratios of LFV
branching ratios. This can be seen easily as follows.  Recall that we
have only three mass parameters, two of which are determined by solar
and atmospheric splittings, while the third is related to the overall
scale of neutrino mass.  The ratio $$\alpha= \Dms/\Dma$$ is well
determined by neutrino oscillation data~\cite{Maltoni:2004ei}.

For the inverse seesaw case we have from eq.(\ref{eigiss}) 
\begin{eqnarray}
\Dma &=&\frac{v_\mu^2}{v_M^4}((a-b)^4-a^4)\\
\Dms &=&\frac{v_\mu^2}{v_M^4}(a^4-(a+b)^4)
\end{eqnarray}
then
\begin{equation}\label{altin}
\alpha=\frac{1-(1-t)^4}{(1+t)^4-1} 
\end{equation}
where $t=-b/a$.  

As mentioned, the main contributions to the LFV processes are those
involving the heavy singlet neutrinos. Then the relevant elements of
the lepton mixing matrix are $K_{ik} \sim M_D\cdot M^{-1}$, and as a
result the $G$ matrix of eq. (\ref{def:G}) is characterized by only
two parameters,
\begin{equation}
G\sim U_{\omega}^T M_0^T M_0 U_{\omega}
\end{equation}
and for inverse seesaw one finds:
\begin{equation}
G\sim
\left(\begin{array}{ccc}
a^2 +\frac{4ab}{3}+\frac{2b^2}{3}& -\frac{1}{3}b(2a+b)&-\frac{1}{3}b(2a+b)\\
-\frac{1}{3}b(2a+b)&  \frac{1}{3}b(4a-b) & a^2 -\frac{2ab}{3}+\frac{2b^2}{3} \\
-\frac{1}{3}b(2a+b)  & a^2 -\frac{2ab}{3}+\frac{2b^2}{3} & \frac{1}{3}b(4a-b)  
\end{array}\right), \nn
\end{equation}
Taking ratios of branching ratios, prefactors cancel and one finds, for 
example for
\begin{equation}\label{brin}
\frac{Br(\tau\to\mu\gamma)}{Br(\tau\to e \gamma)}=
\left(\frac{3+2 t +2 t^2}{2t -t^2}\right)^2,
\end{equation}
where $t$ is the solution of the eq.\,(\ref{altin})

\begin{center}
\begin{figure}[!h]
\includegraphics[angle=0,height=6cm,width=0.48\textwidth]{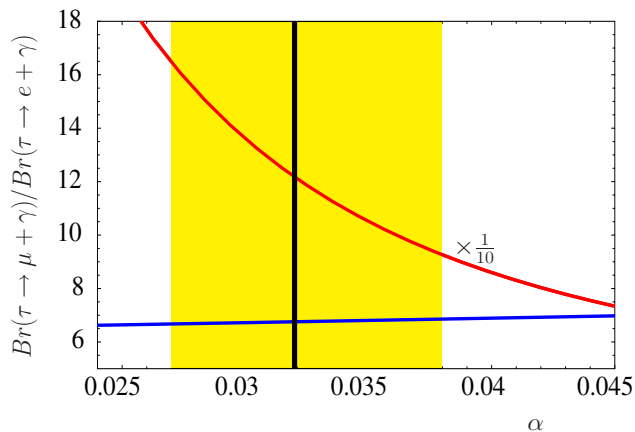}
\caption{$Br(\tau\to\mu\gamma)/Br(\tau\to e \gamma)$ vs $\alpha$ for
  inverse seesaw (red) and linear seesaw (blue). The vertical line
  indicates the best fit value for $\alpha$, the band is the allowed
  $3 \sigma\, \text{C.L.}$ range~\cite{Maltoni:2004ei}.}
  \label{fig3}
\end{figure}
\end{center}

A similar procedure can be carried out for the linear seesaw, using 
eq.(\ref{eiglin}) for the light neutrino mass eigenvalues. One finds
\begin{equation}\label{brl}
\frac{Br(\tau\to\mu\gamma)}{Br(\tau\to e \gamma)}=
\left(\frac{3+u}{u}\right)^2~,
\end{equation}
where $u$ is the solution of the equation
\begin{equation}
\frac{1-(1-u)^2}{(1+u)^2-1}=\alpha~.
\end{equation}
As a result of Eqs.~(\ref{brin}) and (\ref{brl}) we obtain the
predictions illustrated in Fig.~\ref{fig3}. Note the different 
dependence on $\alpha$.

A basic symmetry property of the matrix $G_{ij}$ is mu-tau symmetry,
which implies that $G_{31}=G_{21}$, so that
\begin{equation}
\frac{Br(\tau\to e \gamma)}{Br(\mu\to e \gamma)}=
\left(\frac{m_\tau}{m_\mu}\right)^5\frac{\Gamma_\mu}{\Gamma_\tau}\approx 0.18,
\end{equation}
for both linear and inverse seesaw.  Given the current bounds on
\(\mu\to e \gamma\) we have B(\(\tau \to e \gamma\)) $\lsim 2\times
10^{-12}$ placing a tremendous challenge for the search for lepton
flavor violating tau decays
for testing the prediction given in Fig.~\ref{fig3}.

Before closing this section let us also mention that the TeV neutral
heavy leptons are potentially accessible directly in accelerator
experiments, see, for example, Ref.~\cite{Dittmar:1989yg}.

\section{Discussion}

The inverse and linear seesaw mechanisms provide very interesting
alternative scenarios to the type-I seesaw since the scale of the new
fermions leading to neutrino mass can lie at the TeV scale,
potentially accessible at the LHC.  In this paper we have given two
models based on the $A_4$ discrete flavor symmetry and realizing the
successful tri-bimaximal ansatz for lepton mixing. We have introduced
several Higgs doublets transforming as triplet and singlet
representations of $A_4$. We have assumed that $A_4$ is spontaneously
broken to $Z_3$ in the charged lepton sector and into $Z_2$ in the
neutrino sector which yields the TBM lepton mixing pattern.
Both models are highly predictive as they effectively depend only on
three mass parameters and one Majorana phase, implying no CP violation
in neutrino oscillations.
In contrast to the inverse seesaw, the linear seesaw leads to a lower
bound for neutrinoless double beta decay.

Among their other phenomenological features, the mixing of heavy
neutrinos in the charged electroweak current leads to various lepton
flavor violating decays such as \(l_i \to l_j \gamma\) and $l_i \to
l_j l_k l_k$. In contrast to standard type-I seesaw, here these rates
can be sizeable even in the absence of supersymmetry. Moreover, the
TBM mixing pattern leads to specific predictions for LFV decays as
illustrated, for example, in Fig.~\ref{fig3}. However, within our
particular $A_4$ symmetry realizations, the TBM pattern also implies
that B(\(\tau \to \mu \gamma\)) $\lsim 3\times 10^{-10}$, well below
current experimental sensitivities.

As a final comment, we have only described in this paper results that
follow from exact symmetry realizations of the tri-bimaximal mixing
pattern. It is possible, however, that the symmetry leading to TBM
holds only at some high unification scale and deviations are induced.
Possible radiative effects have been considered for example, in the
framework of supergravity models in Ref.~\cite{Hirsch:2006je}.
For example, in the presence of supersymmetry, broken by soft breaking
terms that do not respect our flavor symmetry, one would have
potentially important corrections that might enhance tau-violating
processes with respect to the predictions presented here.
Finally, let us also mention that, as already noted in
\cite{Malinsky:2005bi} generic inverse and linear seesaw models may be
embedded in an $SO(10)$ framework.  The non-abelian flavor structure
may be incorporated in these models in order to generate the TBM
pattern discussed here, along the lines considered in
Refs.~\cite{Morisi:2007ft} and \cite{Bazzocchi:2008sp}. These are
issues that we hope to take up elsewhere.

\section*{Acknowledgments}

This work was supported by the Spanish grant FPA2008-00319/FPA.


\renewcommand{\baselinestretch}{1}

\end{document}